\documentclass[aps,prx,superscriptaddress,twocolumn]{revtex4-2}
\usepackage[utf8]{inputenc}
\usepackage{bm}
\usepackage{bbm}
\usepackage{bbold}
\usepackage{amsfonts}
\usepackage{amsmath}
\usepackage{amssymb}
\usepackage{graphicx}
\usepackage{mathtools}
\usepackage{upgreek}
\usepackage{complexity}
\usepackage{xfrac}
\usepackage{soul}
\usepackage{xcolor}
\usepackage{dsfont}

\usepackage{url}

\usepackage{txfonts}
\DeclareSymbolFont{matha}{OML}{txmi}{m}{it}
\DeclareMathSymbol{\varv}{\mathord}{matha}{118}

\newtheorem{theorem}{Conjecture}

\begin{document}

\title{Geometric quantum machine learning of BQP$^A$ protocols and latent graph classifiers}

\author{Chukwudubem Umeano}
\affiliation{Department of Physics and Astronomy, University of Exeter, Stocker Road, Exeter EX4 4QL, United Kingdom}

\author{Vincent E. Elfving}
\affiliation{PASQAL, 7 Rue Léonard de Vinci, 91300 Massy, France}

\author{Oleksandr Kyriienko}
\affiliation{Department of Physics and Astronomy, University of Exeter, Stocker Road, Exeter EX4 4QL, United Kingdom}
\affiliation{PASQAL, 7 Rue Léonard de Vinci, 91300 Massy, France}

\date{\today}

\begin{abstract}
Geometric quantum machine learning (GQML) aims to embed problem symmetries for learning efficient solving protocols. However, the question remains if (G)QML can be routinely used for constructing protocols with an exponential separation from classical analogs. In this Letter we consider Simon's problem for learning properties of Boolean functions, and show that this can be related to an unsupervised circuit classification problem. Using the workflow of geometric QML, we learn from first principles Simon's algorithm, thus discovering an example of $\BQP^A \neq \BPP$ protocol with respect to some dataset (oracle $A$). Our key findings include the development of an equivariant feature map for embedding Boolean functions, based on twirling with respect to identified bitflip and permutational symmetries, and measurement based on invariant observables with a sampling advantage. The proposed workflow points to the importance of data embeddings and classical post-processing, while keeping the variational circuit as a trivial identity operator. Next, developing the intuition for the function learning, we visualize instances as directed computational hypergraphs, and observe that the GQML protocol can access their global topological features for distinguishing bijective and surjective functions. Finally, we discuss the prospects for learning other $\BQP^{A}$-type protocols, and conjecture that this depends on the ability of simplifying embeddings-based oracles $A$ applied as a linear combination of unitaries.
\end{abstract}

\maketitle


\textit{Introduction.---}Quantum machine learning (QML) has leaped forward in recent years \cite{Biamonte2017,Benedetti2019rev,Schuld2022PRXQ}, and transformed into a distinct approach to data processing. Typically based on variational approaches \cite{Cerezo2021rev,Tilly_2022}, it draws power from embedding data into quantum states and performing adaptive measurements \cite{Schuld2019feature,schuld2021supervised,Goto2021PRL}. QML protocols include examples of various supervised and unsupervised approaches with examples encompassing classification \cite{Cong2019,PerezSalinas2020datareuploading,SamuelChen2021,SamuelChen2022,zapletal2023errortolerant,Monaco2023,umeano2023learn}, graph-based problems \cite{LPHenry2021,Albrecht2023}, reinforcement learning \cite{Saggio2021}, scientific machine learning \cite{kyriienko2021solving,lubasch2020variational,heim2021quantum,kyriienko2021generalized,Markidis2022qpinns,paine2023quantum}, learning from experiments \cite{Huang2022}, anomaly detection \cite{Herr2021,kyriienko2022unsupervised,wozniak2023quantum,Bermot2023}, and generative modelling \cite{Zoufal2019,Coyle2020,Paine2021, kyriienko2022protocols,kasture2022protocols}. Drivers for using the QML-based data processing go beyond computational speed-up \cite{Schuld2022PRXQ}, and include potential advantages in sampling \cite{Arute2019,Coyle2020}, generalization \cite{Caro2022,gilfuster2023understanding}, data-frugal learning \cite{Cong2019}, and even communication \cite{gilboa2023exponential}.  The \emph{in silico} performance of QML protocols depends on the choice of quantum feature maps for data embedding \cite{SchuldSweke2021PRA,williams2023quantum,umeano2023learn}, the choice of variational ansatz (model expressivity) \cite{Abbas2021}, and scaling of gradients that define trainability \cite{mcclean2018barren,Cerezo2021NatComm,holmes2022connecting,fontana2023adjoint,ragone2023unified}. Various hardware realizations have proven these concepts \cite{JWPan2021,Rudolph2022,glick2022covariant,Herrmann2022,huang2022quantum,Pan2023,Albrecht2023} while revealing challenges with respect to noise and significant requirements for the measurement budget (shots). The community continues to learn what is required for successful deployment of QML protocols.

Contemporary advances of machine learning correspond to building models tailored to considered problems, as defined by their symmetries \cite{bronstein2021geometric}. This concept was successfully applied to QML, leading to the birth of geometric quantum machine learning (GQML) \cite{ZhengStrelchuk2023,JJMeyer2023PRXQ,Larocca2022PRXQ,nguyen2022theory}, also known as group-invariant QML. Based on principles of model invariance and ansatz equivariance, GQML shows promise for training models that are ``just-right'' for the problem \cite{bowles2023contextuality}, and thus boosting their trainability \cite{holmes2022connecting}, as now seen via ansatze based on bounded dynamical Lie algebras \cite{larocca2023theory}. One of the drawbacks is that symmetries may eventually dequantize protocols (enable classical execution) \cite{goh2023liealgebraic}. Recently this was presented as a danger to QML-based approaches \cite{Cerezo2023CSIM}, and the quest for finding QML protocols with exponential advantage is open.
\begin{figure}[t!]
\includegraphics[width=1.0\linewidth]{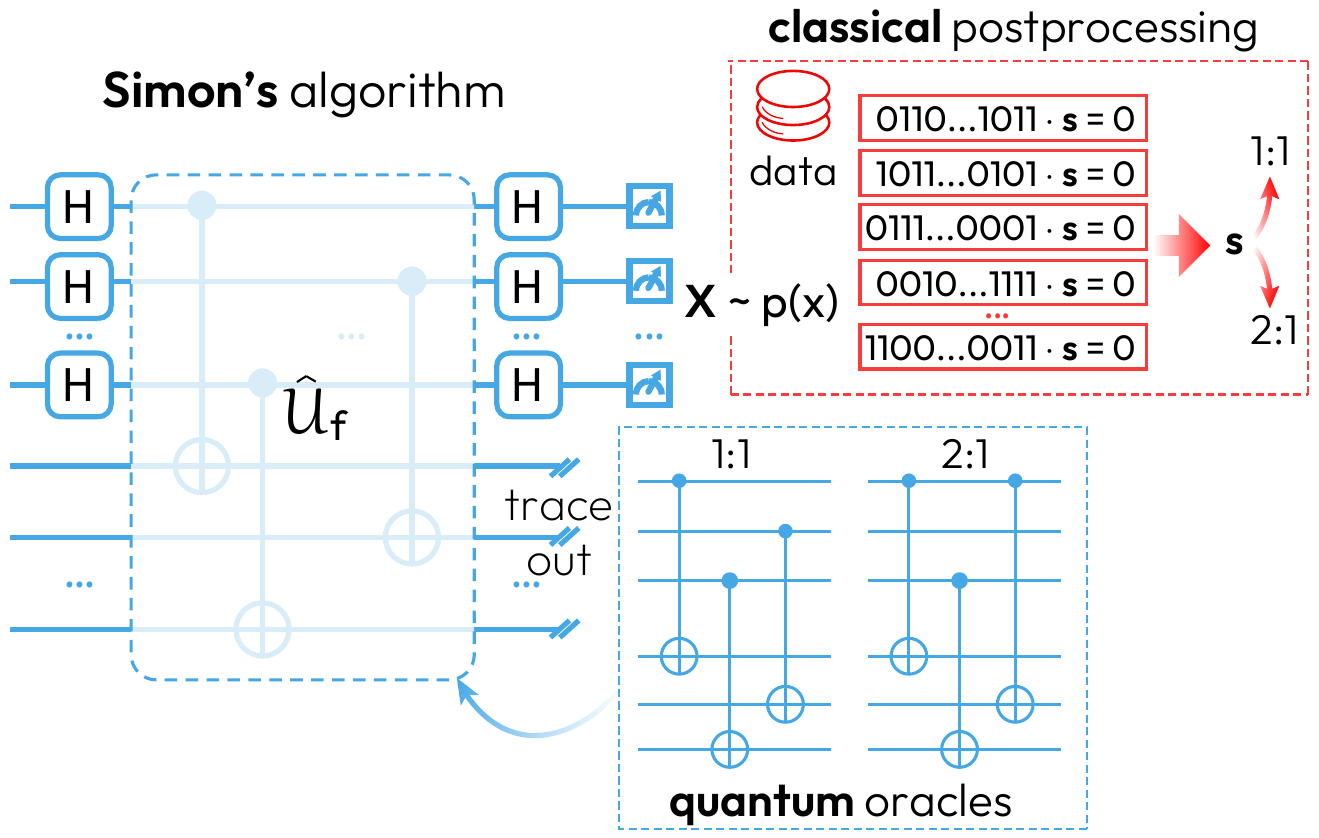}
    \caption{Simon's algorithm based on black-box quantum oracles $\hat{\mathcal{U}}_f$ that encode $N$-bit Boolean functions over $2N$ quantum registers. Due to quantum parallelism the measured samples $X \sim p(x)$ contain information about a hidden bitstring $s$. This is obtained from classical post-processing, and defines whether the function is 1:1 or 2:1.}
    \label{fig:Simons}
\end{figure}

Stepping back from QML to textbook quantum computing protocols (Shor's, Simon's etc. \cite{nielsen2010quantum,Shor1997,Simon1997}), we may wonder if the ability of quantum computers in solving Abelian hidden subgroup problems via Shor-type algorithms can inspire the development of provably-efficient QML programs. One attempt showed this to be possible when using kernel-based approaches with embeddings based on Shor's circuit itself \cite{Liu2021NatPhys}. Other approaches used variational circuits to recover unitaries for different algorithms including Grover's, Simon's and state overlap circuits~\cite{Morales2018,wan2018learning,Cincio2018}, targeting the search of optimal basis transformation. It remains unclear how to use QML in a scalable way, and generally exploit \emph{learning} advantages. We need a systematic approach for discovering new algorithms in the $\BQP$ family, starting from recovering known cases.
\begin{figure*}[t!]
\includegraphics[width=1.0\linewidth]{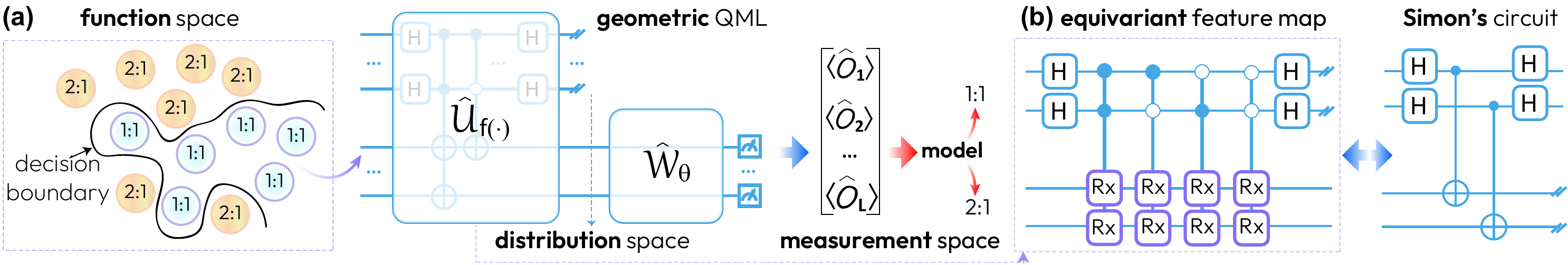}
    \caption{\textbf{(a)} Workflow for 2:1 and 1:1 function classification based on GQML, where data are mapped from the function to quantum space by equivariant embedding, and probability distributions are studied based on invariant observables. This is equivalent to Simon's algorithm \textbf{(b)}.}
    \label{fig:GQML}
\end{figure*}

In this Letter we develop a GQML-based approach for function learning, and apply it to Simon's problem. We are able to learn the exponentially fast Simon's algorithm \cite{Simon1997}, being a showcase of $\BQP^A$ protocol that separates quantum algorithms from classical probabilistic analogs in $\BPP$. The crucial development is an equivariant feature map for function loading and post-processing of results, while for this specific problem the learnt ansatz is trivial. Visualizing the data as computational hypergraphs, we connect the process with learning topological properties of exponentially large latent-space graphs. While the utility of the provided example remains limited due to black-box operation inherent to Simon's algorithm, we conjecture that there may be problems that avoid this regime yet enjoy the exponential speed-up, conditioned on properties of the data.


\textit{Background: textbook Simon's algorithm.---}We start by describing a specific problem that we intend to solve to showcase the capabilities of GQML. This corresponds to Simon's problem of determining properties of Boolean functions $f$ of an $N$-bit argument $x \in \mathcal{X} = \{0,1\}^N$, with $f: x \mapsto f(x) \in \mathcal{X}$ \cite{nielsen2010quantum,Simon1997}. Specifically, the goal is to determine whether $f$ is a bijective one-to-one (1:1) function that maps each input string to a unique output string, or a surjective two-to-one (2:1) function that admits the same output for two different inputs. The latter property can be associated to the existence of a \emph{hidden} bitstring $s \in \mathcal{X}$ such that for two inputs $x, y \in \mathcal{X}$ we have $f(x) = f(y) = f(x \oplus s)$, implying that bitwise XOR $x \oplus y = s$. For $f_{2:1}$ instances we have non-zero hidden strings $s$, while $f_{1:1}$ correspond to $s = 0^N$. The corresponding decision problem is distinguishing $f_{1:1}$ and $f_{2:1}$ instances. Simon's quantum algorithm is able to find $s$ via repeated queries to a \emph{black-box} $\mathcal{U}_f$, implemented as a quantum circuit. Importantly, Simon's problem falls into the category of Abelian hidden subgroup problems, and is the first example to show the oracle separation between $\BQP$ and $\BPP$ classes \cite{nielsen2010quantum}. Namely, for a particularly chosen oracle $A = \mathcal{U}_f$ one can develop a quantum algorithm requiring exponentially fewer oracle queries as compared to classical probabilistic approaches, meaning $\BQP^{\mathcal{U}_f} \neq \BPP$.

Simon's algorithm is summarized in Fig.~\ref{fig:Simons}. It assumes access to the unitary $\hat{\mathcal{U}}_f$ that encodes function $f$, implemented on the $2N$-qubit register as $\hat{\mathcal{U}}_f|x\rangle|y\rangle = |x\rangle |f(x)\oplus y\rangle$. While generally we cannot access the oracle (black-box nature prevents it), in practice one can build them based on CNOTs that connect the top and bottom registers (Fig.~\ref{fig:Simons}, blue dashed box). The oracles for 1:1 functions have CNOTs placed with unique non-repeated pairs, while 2:1 oracles include multiple CNOTs controlled by the same qubits. When the top register of $n$ qubits is prepared in equal superposition (layer of Hadamards in Fig.~\ref{fig:Simons}), the state is transformed into an effective superposition of outcomes $2^{-N/2} \sum_{x \in \mathcal{X}} |x\rangle|f(x)\rangle$. Upon applying another layer of Hadamards, tracing out the bottom register and measuring the top register, we get sample bitstrings $X \sim p(x)$ coming from a probability distribution that depends on $s$. For 1:1 functions this distribution is flat, $p_{1:1}(x) = 1/2^N$, while for 2:1 functions it is colored, as only half of the $2^N$ possible bitstrings are present in the distribution. The crucial part of Simon's algorithm is the \emph{classical post-processing}, which relies on drawing samples and solving a system of equations for the binary variable $s$ (Fig.~\ref{fig:Simons}, red dashed box). In this work our aim is to learn the algorithm from the basic principles of geometric quantum machine learning, leading to the same predictions.


\textit{Background: geometric QML.---}We begin with the formalism for geometric quantum machine learning that has been detailed in several studies~\cite{JJMeyer2023PRXQ,Larocca2022PRXQ,nguyen2022theory,ZhengStrelchuk2023,bowles2023contextuality}. 
Let us assume we have a classical dataset of $M$ samples, $\mathcal{D} = \{ \bm{x}_m, y_m\}_{m=1}^{M}$, where $x_m$ belong to an input domain $\mathcal{X}$ and $y_m$ belong to a label domain $\mathcal{Y}$. The dataset assumes an underlying function $g: \mathcal{X} \rightarrow \mathcal{Y}$ such that $g(x_m)=y_m$. Our aim is to train a parameterized quantum model $h_\theta$ to approximate this target function across all inputs. We define a quantum model as
\begin{equation}
    h_\theta (x)=\langle\psi_0|\hat{U}^\dagger(x)\hat{W}^\dagger(\theta)\hat{\mathcal{O}}\hat{W}(\theta)\hat{U}(x)|\psi_0\rangle,
\end{equation}
with $|\psi_0\rangle$, $\hat{U}(x)$, $\hat{W}(\theta)$, $\hat{\mathcal{O}}$ representing the initial state, feature map, adjustable ansatz and measured observable, respectively.

The core concept of geometric (group-invariant) machine learning is in introducing models that obey symmetries and ensure the corresponding label invariance. This can be visualized in a simple form based on the paradigmatic image classification example: a picture of a cat which has been rotated or had its pixels translated still remains an image of a cat, thanks to the rotational and translational invariance. Suppose we have a symmetry group $\mathcal{S}$ with a corresponding representation $V_\sigma: \mathcal{S} \rightarrow Aut(\mathcal{X})$ \cite{JJMeyer2023PRXQ,nguyen2022theory}, we say that the function $g$ is invariant under the action of $\mathcal{S}$ if 
\begin{equation}
    g(V_\sigma[x])=g(x) \quad \forall x \in \mathcal{X}, \sigma \in \mathcal{S}.
\end{equation}
Since this is a property of our target function, it is desirable to build a model $h_\theta$ such that it is \emph{invariant} under symmetry: $h_\theta(V_\sigma[x])=h_\theta(x)~\forall ~\theta,~x \in \mathcal{X}, \sigma\in\mathcal{S}$. We note that for each symmetry $\sigma \in \mathcal{S}$, we have a corresponding unitary representation $\hat{U}_\sigma$ which acts on the Hilbert space. Several ingredients can be combined to produce an invariant model \cite{JJMeyer2023PRXQ} based on: 1) \emph{invariant} initial state, $\hat{U}_\sigma|\psi_0\rangle = |\psi_0\rangle$; 2) \emph{equivariant} embedding, $\hat{U}(V_\sigma[x])=\hat{U}_\sigma\hat{U}(x)\hat{U}_\sigma^\dagger$; 3) \emph{equivariant} ansatz, $[\hat{W}(\theta), \hat{U}_\sigma] = 0$; and 4) \emph{invariant} measured observable satisfying $\hat{U}_\sigma\hat{\mathcal{O}}\hat{U}_\sigma^\dagger = \hat{\mathcal{O}}$. 
Once we know the symmetries of the problem, the aim is to encode them into the GQML workflow, starting from data loading and ending with symmetry-preserving measurement.


\textit{Results: learning Simon's algorithm with GQML and equivariant feature maps for distribution loading.---}Returning back to Simon's problem, we ask a question: can we see it as a machine learning task? The answer is affirmative, since learning hidden strings $s$ can be seen as a regression problem (admittedly, with the difference of being a supervised task). Simon's decision problem can be set up as a function classification problem, and approached by unsupervised learning protocols.
An important distinction of Simon's problem as compared to other machine learning problems is the type of data, as we classify $f_{1:1}$ and $f_{2:1}$ Boolean functions assigning labels $y_{1:1} = 0$ and $y_{2:1} = 1$. The samples of the corresponding dataset $\mathcal{D}$ are no longer scalars, $\mathcal{D} = \{ f_m(\cdot), y_m\}_{m=1}^{M}$, with $f_m(\cdot)$ belonging to the set of functions $\mathcal{F}$ that retain their properties no matter which input string we choose. Therefore our target function $g$ now becomes a functional, $g(f_m) = y_m$.
%
We emphasise that the function $f_m$ has a label $y_m$ which does not depend on the $x$ input, $g(f_m(x')) = g(f_m(x)) = y_m$. In essence, $g$ has an invariance with respect to the function inputs $x$. For a function defined on $N$ bits there are $2^N$ possible inputs, and we can define a permutation-type symmetry $\mathcal{S}_{2^N}$ with corresponding representations $V_{x,x'}$ such that $f_m(V_{x,x'}[x]) = f_m(x')$. This is the ``symmetry'' we must encode into our quantum model.

To do this, we make use of the \emph{twirling} method \cite{JJMeyer2023PRXQ,nguyen2022theory}, commonly used to determine the gateset required to build an equivariant ansatz. Given a particular gateset $\mathcal{G}$, we can build an equivariant gateset $\mathcal{T}_{\hat{U}}[\mathcal{G}]=\{\mathcal{T}_{\hat{U}}[\hat{G}] | \hat{G} \in \mathcal{G}\}$, where $T_{\hat{U}}[\hat{G}] = \frac{1}{|\mathcal{S}|}\sum_{\sigma\in\mathcal{S}}\hat{U}_\sigma \hat{G} \hat{U}_\sigma^\dagger$. Each of these twirled operators commutes with each unitary representation of the symmetry group, $[T_{\hat{U}}[\hat{G}],\hat{U}_\sigma] = 0~\forall~\hat{G} \in \mathcal{G},~\sigma \in \mathcal{S}$, hence an ansatz built from these operators is equivariant. 
We stress that here twirling is applied to the \emph{embedding} itself. For a particular function $f_m$, we have a set of unitaries $\{\hat{U}(f_m,x) | x\in\{0,1\}^{\otimes N} \}$ which encode the function evaluated at each input $x$, $\hat{U}(f_m,x)|\phi\rangle = |f_m(x)\rangle$, with $|\phi\rangle = |0\rangle^{\otimes N}$. We then apply twirling to an initial state $\rho_0 = |\phi\rangle\langle\phi|$ to build the embedding
\begin{equation}\label{twirling}
    \rho(f_m) = \frac{1}{2^N}\sum_{x\in\{0,1\}^{\otimes N}}\hat{U}(f_m,x)\rho_0\hat{U}^\dagger(f_m,x).
\end{equation}
We note that we only need to twirl with the set of $2^N$ unitaries corresponding to the representations $V_{0,x}$. The unitaries performing the mapping $|f_m(x)\rangle \rightarrow |f_m(x')\rangle$ can be formed from combinations of this set.

In practice, we apply the twirling to the embedding by applying a linear combination of unitaries (LCU) \cite{childs2012hamiltonian,childs2017quantum,Childs2020} on the initial state. Each of the unitaries $\hat{U}(f_m,x)$ is applied one by one as controlled operators in an extended Hilbert space. From our definition we can see that each unitary needs to map from the zero state to a computational basis state. This can be done with a set of bitflips, i.e. $\hat{U}(f_m,x) \in \{\hat{X},\mathbbm{1}\}^{\otimes N}$, or equivalently up to a phase, $\hat{U}(f_m,x) \in \{R_x(\pi),R_x(2\pi)\}^{\otimes N}$ (Fig.~\ref{fig:GQML}(b)). 
After each circuit run the ancilla qubits are measured, and if the outcome is $0^N$, we have successfully applied the LCU, getting  $|\psi(f_m)\rangle = \frac{1}{2^N}\sum_{x \in \{0,1\}^{\otimes N}} \hat{U}(f_m,x)|\phi\rangle$ \cite{williams2023quantum}. However, if we trace out this ancilla register, we recover the density state embedding $\rho(f_m)$. It is key to note that this density operator is \emph{exactly the same} as the one obtained from running the corresponding Simon's circuit and tracing out the bottom $N$ qubits (Fig.~\ref{fig:GQML}(b)). This operator represents a mixed state encoding the probability of measuring each function output. 

The symmetry described previously is on the level of the function \emph{inputs}. However, we also have symmetries on the level of the function \emph{outputs}. To specify, $g(V_{\mathrm{bp}}[f_m]) = g(f_k) = g(f_m)$, where $V_{\mathrm{bp}}$ is the representation of the group of \emph{bitflips} and \emph{permutations}, $S_{2^N}\rtimes Z_2^{2^N}$ \cite{JJMeyer2023PRXQ}. 
%
On the level of the Hilbert space, we can see that our embedding $\rho(f_m)$ is also equivariant with respect to both bitflips and permutations:
$\rho\left(V_{\mathrm{bp}}[f_m]\right) = \rho(f_k)=\hat{U}_{\mathrm{bp}}\rho(f_m)\hat{U}^\dagger_{\mathrm{bp}}$, with the corresponding unitary representations $\hat{U}_{\mathrm{bp}} \in \{\mathrm{SWAP}_{ij}, \hat{X}_i ~\forall i,j\in \{1,2,\ldots,N\}\}$.


%
We now turn to the choice of ansatz. We apply twirling over the set of symmetry representations $\{\hat{U}_{\mathrm{bp}}\}$ to find generators to build an equivariant ansatz. However, this symmetry group restricts the available gateset to a set of $X$ rotations, which have a trivial effect on the embedding. Our model is already in the correct subspace, so we choose $\hat{W}(\theta) = \mathbbm{1}$, $\theta = \emptyset$, and then make measurements directly on $\rho(f_m)$ to define our model.

The final step is the choice of observable $\hat{\mathcal{O}}$. Ideally we require an observable which will allow us to distinguish between $f_{1:1}$ and $f_{2:1}$, while maintaining model invariance. We find that a collection of $\hat{Z}$ operators satisfies both conditions,
\begin{equation}\label{Cost}
    \hat{\mathcal{O}} = \sum_i Z_i + \sum_{i<j} Z_iZ_j + \sum_{i<j<k} Z_iZ_jZ_k + \ldots + Z_1Z_2\ldots Z_n.
\end{equation}
For 1:1 functions we find that $\mathrm{tr}\left(\rho(f_m)\hat{\mathcal{O}}\right) = 0$, whereas $\mathrm{tr}\left(\rho(f_m)\hat{\mathcal{O}}\right) = 1$ for 2:1 functions. This separation is maintained under the bitflip and permutation symmetry transformations, $\mathrm{tr}\left(\hat{U}_{\mathrm{bp}}\rho(f_m)\hat{U}^\dagger_{\mathrm{bp}}\hat{\mathcal{O}}\right) = \pm \mathrm{tr}\left(\rho(f_m)\hat{\mathcal{O}}\right)$.


\textit{Results: clustering and unsupervised learning of Boolean functions.---}We proceed to perform numerical simulations to test this approach. Choosing a system size $N = 6$, we generate a dataset $\mathcal{D} = \{ f_m, y_m\}_{m=1}^{M}$ with $M=120$ unique functions, ensuring that half of the functions are 1:1 $(y_m=0)$, and the other half, 2:1 $(y_m=1)$. Each function is loaded via the GQML process detailed previously, creating a density operator $\rho(f_m)$ \eqref{twirling}. We then measure the cost operator $\hat{\mathcal{O}}$ \eqref{Cost} on this mixed state using a fixed number of shots, $N_s = 5000$. We take the mean and variance of these measurements as our two features for classification: feature vector $v_{f_m} = \left(\langle\hat{\mathcal{O}}\rangle_{f_m}, \mathrm{var}(\hat{\mathcal{O}})_{f_m}\right)$.
\begin{figure}[ht!]
\includegraphics[width=1.0\linewidth]{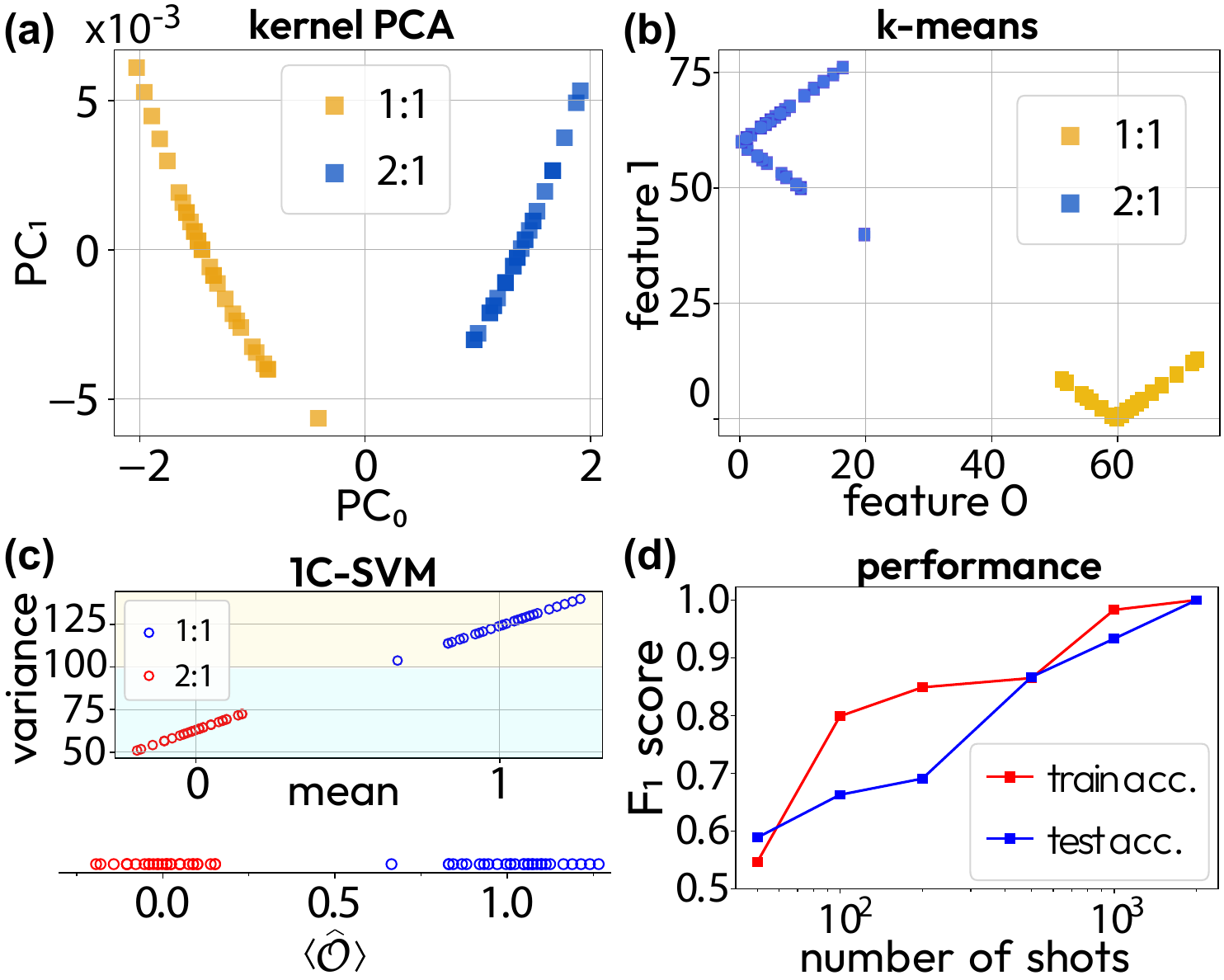}
    \caption{Results for the classical post-processing of equivariantly embedded functions. We used unsupervised approaches to classification corresponding to kernel PCA \textbf{(a)}; k-means clustering \textbf{(b)}; and one-class SVM with linear kernels \textbf{(c)}. In \textbf{(d)} we show how the classification performance (F$_1$ score) increases with the number of shots.}
    \label{fig:SVM}
\end{figure}
We then continue by applying classical unsupervised machine learning methods to the data, and the results are shown in Fig. \ref{fig:SVM}. We first use two clustering methods: kernel Principle Component Analysis (PCA) \cite{Scholkopf1997} and K-Means \cite{Lloyd1982,Forgy1965ClusterAO}, as implemented in \textsf{scikit-learn} \cite{scikit-learn}. We can see that the separation between the two classes of functions is clear, showing the success of this hybrid quantum-classical procedure.

Next, we approached the problem from the anomaly detection perspective \cite{kyriienko2022unsupervised}. One-class Support Vector Machine (SVM) \cite{Scholkopf1999} is an unsupervised model which learns on data from one class only. Therefore, we now split the data equally into a training and test dataset, each set containing thirty 1:1 and 2:1 instances. We train the SVM on the training data from the 1:1 class only, before applying it to the test data containing unseen samples from both classes. In this way, we test the SVM's ability to identify unseen 1:1 functions, as well as distinguish the anomalous 2:1 data. 

Fig. \ref{fig:SVM}(c) shows the results for this classification. In the top subplot, the boundary between the two classes is indicated by the blue and yellow regions. We see that with a sufficient number of shots taken to acquire the feature vectors $v_{f_m}$, the SVM can identify 1:1 functions and distinguish them from the 2:1 functions successfully. The separation is apparent even in one dimension; as the bottom subplot in this figure shows, the two classes of functions can be distinguished solely from the mean of the measurements of $\hat{\mathcal{O}}$. 
\begin{figure*}[t!]
\includegraphics[width=1.0\linewidth]{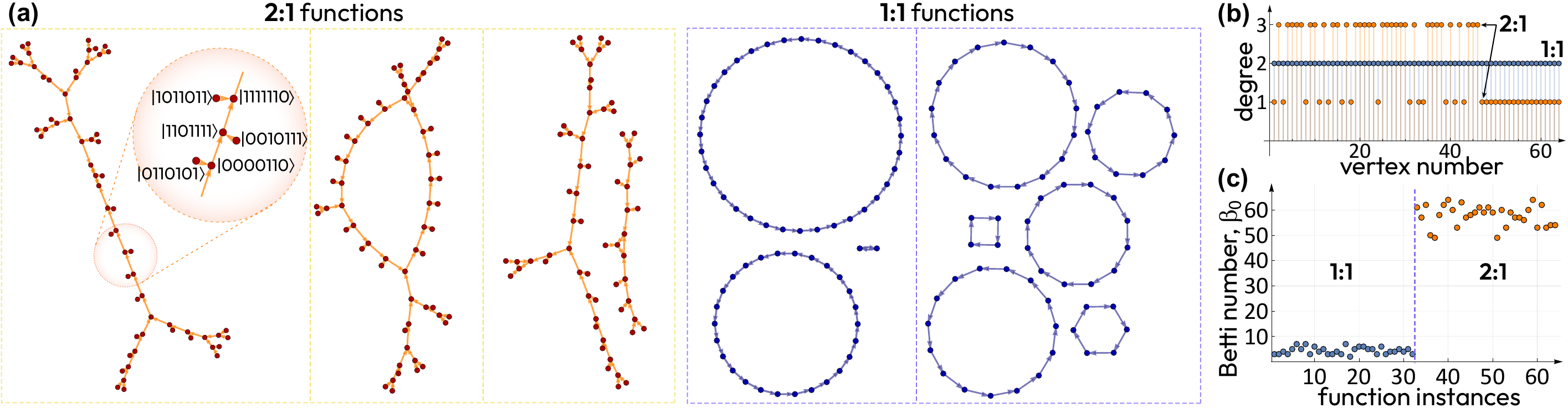}
    \caption{Visualization of function learning process as a latent-space graph classification. \textbf{(a)} Qualitatively different directed computational hypergraphs that emerge from 2:1 and 1:1 functions. Each vertex is a basis state ($N=6$), and directed edges induced by the action of oracle $\hat{U}_f$ point to output states. \textbf{(b)} Degree shown for every vertex in the directed hypergraph. For $f_{1:1}$ functions this is flat, while for $f_{2:1}$ it is non-uniform. \textbf{(c)} Graph Betti number of the zeroth order, $\beta_0$, showing that classification can be performed based on topological properties.}
    \label{fig:graphs}
\end{figure*}
These figures were generated using $N_s = 5000$ shots for each function to emphasise the large separation between the two classes of functions. However, this number of shots is not necessary for high accuracy classification. To show this, we performed the same procedure using one-class SVM, but we varied the number of shots used to generate the feature vectors $v_{f_m}$. We then measured the $F_1$ score of the model acting on the training and test datasets, a measure of predictive performance \cite{Sasaki2007}. Our results are shown in Fig.~\ref{fig:SVM}(d). While with too few shots distinguishing anomalies is challenging, increasing the shot number improves performance rapidly. Since post-processing includes efficient strategies like in the original Simon's algorithm \cite{Simon1997}, more advanced classical networks can also be learnt.


\textit{Visualization: function learning as topological graph recognition.---}So far we have addressed the problem as a ``dry'' function learning task, and formally built a symmetry-enabled GQML framework for distinguishing quantum circuits that represent the functions. Can we visualize this process in a way that provides extra understanding of the learning process? We believe the answer is yes, and approach this with a specific problem visualization. A tool we use is a directed hypercube representation of Boolean functions \cite{CLARKE1994}. We represent each vertex as a bitstring (basis state). An action of a $\hat{U}_f$ represents a directed edge, which connects an input $|x_i\rangle$ state with a final state $|x_f\rangle = \hat{U}_f |x_i\rangle$. The connectivity of this computational hypergraph can then reveal properties of functions.

We visualize examples of directed hypergraphs for 2:1 and 1:1 functions in Fig.~\ref{fig:graphs}(a). In the former case the typical shape is of \emph{axon}-type \cite{Rabinovich2006}, and features numerous branches (can be seen as \emph{dendrites} in the magnifier in Fig.~\ref{fig:graphs}a, left orange box). This is a result of the 2:1 nature of the function, imposing two outputs from the same input at small scale (\emph{local} structure), and thus also defining the \emph{global} structure. From the 1:1 function visualization in Fig.~\ref{fig:graphs}a (blue boxes) we observe that corresponding hypergraphs have a cyclic structure, stemming from the bijective nature of generating functions. The global structure corresponds to directed loops of varying length.

We proceed to analyze properties of the underlying directed hypergraphs. Intriguingly, we find that there are no special graph families that can fully include 2:1 or 1:1 functions (e.g. being a planar, bipartite, tree, Hamiltonian, or acyclic graph), and these attributes are instance-dependent. There is also no clear distinction based on a clustering coefficient or clique sizes. One observation is that at the local level graphs can be distinguished by their vertex degree (Fig.~\ref{fig:graphs}b), where $f_{1:1}$ has the same degree of two for every vertex (single mode zero variance degree distribution), while $f_{2:1}$ has a two-mode degree distribution. This reflects the probability distributions for equivariantly loaded functions $\hat{U}_f$, which is uniform for bijections and anti-concentrated distribution for surjective functions. However, this analysis is highly specific to the chosen problem. Go beyond the local analysis, we note that the global structure difference of 2:1 and 1:1 can be seen in the topological properties, namely the Betti numbers of the graphs \cite{Edelsbrunner_2002,Zomorodian_2004,Lloyd2016,Ubaru2021,scali2023quantum}. We plot the zeroth order Betti number $\beta_0 = |\mathcal{C}_G|$ as a number of connected components $\mathcal{C}_G$ for a directed graph $G$ (as defined in Wolfram Language \cite{connectedgraphcomponents}). The results are plotted in Fig.~\ref{fig:graphs}(c), with the first 32 instances being 1:1 functions, and the rest being 2:1 functions. We can see a clear separation between graphs in terms of the topological feature. The same holds for the first-order Betti number, also known as a cyclomatic number --- linking the quantum circuit analysis to a program complexity analysis. Finally, we can also distinguish graphs \cite{AlbertBarabasi2002} by checking the total length of cycles that is maximal for $f_{1:1}$ and remains small for $f_{2:1}$ instances.

Motivated by the visualization, let us try understanding its implications for function learning and classification. In the classical case each query to (oracle of) $f$ uncovers one part of a graph (being exponentially large in the system size). We can choose to use depth-first or breadth-first approaches, but ultimately the decisions we make are based on the local structure of the directed hypergraph. We can be certain about the class of a function only when we: 1) have prior knowledge on function properties; 2) hit a lucky strike (i.e. reveal two edges coming into the same vertex). On the contrary, the quantum algorithm does have an access to global properties of the hypergraph, albeit in a limited probabilistic manner. By the equivariant LCU-type embedding we generate multiple edges, but require measurements to draw conclusions. One possibility is that quantum approaches effectively exploit the global structure of functions and learn to classify circuits based on topological features of graphs formed in the latent space.


\textit{Discussion: prospects of GQML for $\BQP^A$ problems.---}We have shown one example where the known efficient algorithm for solving a $\BQP$-type problem can be learnt within the geometric quantum machine learning framework. Can this be applied to solve other problems, going beyond Simon's example? Generally, yes, once we supply data in a quantum form and know relevant symmetries. One of the hypothetical targets may include circuits $\hat{U}_f$ within $\BQP$ complete computational complexity class and Promise-$\BQP$. However, we shall be careful with respect to several points.

First, the essence of the GQML approach is the equivariant loading of the function defined by oracle $A$ (or known circuit). Given the input size of $x$ with $2^N$ computational states, the ancillary register requires $N$ qubits (being space efficient), but the depth being in principle exponential when using the LCU approach. For Simon's example we have seen how the LCU is efficiently reduced into an $\mathcal{O}(N)$ circuit that consists of CNOTs only. This leads us to the following conjecture.
\begin{theorem}
Given an oracle $A$ for an $n$-bit function, one can develop an efficient GQML protocol being in $\BQP^A$ for learning its properties if the LCU-based embedding can be simplified into an $\mathcal{O}(\mathrm{poly}(n))$ depth circuit.
\end{theorem}
Similarly, the complexity can be analyzed in terms of space-time tensor network contraction \cite{kasture2022protocols}.

The second point concerns the oracle-based nature of the considered example, where we stress that Simon's problem and the algorithm assume a black box scenario (we do not access the circuit directly). Once the oracle (circuit $\hat{U}_f$) is known, we have enough information to classify underlying functions without even running these circuits. This is similar to the recent hot debates around Grover's protocol operation and utility \cite{stoudenmire2023grovers}. Transitioning into the practical plane, we need to tackle problems with circuits that are not easy to analyze, yet admit the LCU embedding to be compressed (see Conjecture 1 above). One option here is a distributed learning architecture \cite{gilboa2023exponential,kumar2023expressive,Tan2022} or blind quantum computing \cite{BroadbentKashefi2009,Fitzsimons2017,li2023blind,drmota2023verifiable}, where data preparation is performed by one party (no or limited knowledge of the problem is provided) while classification is performed on quantum data being sent to a user site.


\textit{Conclusion.---}In this work we started paving the road between $\ BQP^A$-type protocols offering an exponential advantage (in $A$ oracle-based setting) and quantum machine learning. Using its group-equivariant version enhanced by symmetries, we effectively rediscovered Simon's algorithm for analyzing Boolean functions. This has shown that: 1) solving Simon's decision problem can be interpreted as an unsupervised QML task, corresponding to classifying one-to-one (bijective) functions and two-to-one (surjective) functions; 2) geometric QML models require equivariant data loading and can excel with function-based data; 3) twirling-based equivariant feature maps, represented by LCU feature maps over the function argument domain, are equivalent to introducing quantum parallelism; 4) inherently, function learning can be related to learning latent space topological properties of directed computational hypergraphs. We stress that data embedding and classical post-processing play a huge role, as expected for high-performing QML \cite{Cerezo2023CSIM}, and GQML is also advantageous in the absence of variational training. We conjecture that other $\BQP^A$ protocols can be learnt for other choices of $A$.

\textit{Acknowledgement.---}We acknowledge the funding from UK EPSRC award under the Agreement EP/Y005090/1, and thank PASQAL for the support.




%

\end{document}